Tomasz M. Stawski*[ab], Rogier Besselink[ac], Konstantinos Chatzipanagis[ad], Jörn Hövelmann[ae], Liane G. Benning[af], Alexander E. S. Van Driessche*[c]

# Nucleation Pathway of Calcium Sulfate Hemihydrate (Bassanite) from Solution: implications for Calcium Sulfates on Mars

Tomasz M. Stawski*[a,b], Rogier Besselink[a,c], Konstantinos Chatzipanagis[a,d],
Jörn Hövelmann[a,e], Liane G. Benning[a,f], Alexander E. S. Van Driessche*[c]

\* - corresponding author(s):

tomasz.stawski@bam.de; alexander.van-driessche@univ-grenoble-alpes.fr

[a]German Research Centre for Geosciences, GFZ, Interface Geochemistry, Telegrafenberg, 14473 Potsdam, Germany;

[b]Federal Institute for Materials Research and Testing (BAM), 12205 Berlin, Germany;

[c]Université Grenoble Alpes, Université Savoie Mont Blanc, CNRS, IRD, IFSTTAR, ISTerre, F-38000 Grenoble, France;

[d]Theoretical and Physical Chemistry Institute, National Hellenic Research Foundation, 48 Vassileos Constantinou Ave., 116 35 Athens, Greece;

[e]Remondis Production GmbH, Brunnenstraße 138, 44536, Lünen, Germany;

[f]Department of Earth Sciences, Freie Universität Berlin, 12249 Berlin, Germany.

Orcid:

T.M. Stawski https://orcid.org/0000-0002-0881-5808

R. Besslink https://orcid.org/0000-0002-2027-9403

K. Chatzipanagis https://orcid.org/0000-0002-7043-9364

Jörn Hövelman https://orcid.org/0000-0002-1961-653X

L.G. Benning https://orcid.org/0000-0001-9972-5578

A.E.S. Van Driessche https://orcid.org/0000-0001-8591-487X

**Abstract**

$CaSO_4$ minerals (i.e. gypsum, anhydrite and bassanite) are widespread in natural and industrial environments. During the last several years, a number of studies have revealed that nucleation in the $CaSO_4$-$H_2O$ system is non-classical, where the formation of crystalline phases involves several steps. Based on these recent insights we have formulated a tentative general model for calcium sulfate precipitation from solution. This model involves primary species that are formed through the assembly of multiple $Ca^{2+}$ and $SO_4^{2-}$ ions into nanoclusters. These nanoclusters assemble into poorly ordered (i.e. amorphous) hydrated aggregates, which in turn undergo ordering into coherent crystalline units.

The thermodynamic (meta)stability of any of the three $CaSO_4$ phases is regulated by temperature, pressure and ionic strength with gypsum being the stable form at low temperatures and low to medium ionic strengths, and anhydrite the stable phase at high temperatures and lower temperature at high salinities. Bassanite is metastable across the entire phase diagram but readily forms as the primary phase at high ionic strengths across a wide range of temperatures, and can persist up to several months. Although the physicochemical conditions leading to bassanite formation in aqueous systems are relatively well established, nanoscale insights into the nucleation mechanisms and pathways are still lacking. To fill this gap, and to further improve our general model for calcium sulfate precipitation, we conducted in situ scattering measurements at small- and wide-angles (SAXS/WAXS) and complemented these with in situ Raman spectroscopic characterization. Based on these experiments we show that the process of formation of bassanite from aqueous solutions is very similar to the formation of gypsum: it involves the aggregation of small primary species into larger disordered aggregates, only from which the crystalline phase develops. These data thus confirm our general model of $CaSO_4$ nucleation and provide clues to explain the abundant occurrence of bassanite on the surface of Mars (and not on the surface of Earth).

**Introduction**

$CaSO_4$ minerals (i.e. gypsum, anhydrite and bassanite) are widespread on Earth and Mars, and are involved in numerous high-value industrial processes (e.g. [1,2]). Spurred by both the geological and industrial interests, extensive research efforts have been dedicated to understand the precipitation mechanisms of these minerals under a broad set of conditions. Until recently, however, the details of the first step in the mineral formation process, i.e. nucleation, have remained elusive. With the advent of more powerful observation techniques in the last decade the nucleation of calcium sulfate phases has been explored at the nanoscale (e.g. [3–6]). The general consensus of these studies is that the formation of crystalline phases taking place in the $CaSO_4$-$H_2O$ system involves several steps, including nanosized precursor and intermediate phases. In this multistep nucleation pathway, water activity plays a key role in determining the final crystalline phase[7–9]. Noteworthy, this particle-mediated crystallization route is "fossilized" in the internal structure of the formed crystals[10]. Based on these recent insights we have formulated a tentative general model for calcium sulfate precipitation from solution[1]. Starting from a undersaturated solution, in which calcium and sulfate occur both as individual ions and associated ion pairs[11], nanosized particles are formed, once supersaturation is established, through the assembly of multiple $Ca^{2+}$ and $SO_4^{2-}$ ions into elongated, rather flexible, nanoclusters[4]. When a critical concentration of those primary species is reached, these nanoparticles start to assemble into poorly ordered (i.e. amorphous) hydrated aggregates. In order to transform into ordered arrays of $CaSO_4$ with more (gypsum), less (bassanite) or no (anhydrite) interspersed structural water, the aggregates reorganize themselves through alignment and coalescence of the primary species to form coherent crystalline units while expelling water.

The thermodynamic (meta)stability of any of the three $CaSO_4$ phases is regulated by temperature, pressure and ionic strength considerations with gypsum being the stable phase at low temperatures (<~55°C) and low to medium ionic strengths, while anhydrite is the stable form at higher temperatures (>~55°C) and low ionic strength (and at lower temperatures for high ionic strengths). Bassanite is metastable across the entire phase diagram, but forms as the primary phase at high ionic strengths across a wide range of temperatures, and can persist up to several months[1,7]. Contrary to gypsum , hardly no molecular-level details are available about the early stages of bassanite (and anhydrite) formation from aqueous solutions.

However, bassanite precipitation induction times were measured as a function of supersaturation and temperature, from which the interfacial free energy (~9 mJ/m$^2$)[12] of bassanite nucleated in concentrated electrolyte solutions could be extracted. Other studies focused on establishing the temperature and salinity regions where bassanite is the main phase formed from direct precipitation experiments[7,13,14]. Interestingly, the temperature for the primary precipitation of bassanite can be significantly lowered by increasing the ionic strength of the reacting solutions, in particular by using highly concentrated electrolyte solutions[7,13,14]. For example, bassanite is formed already at 80°C in the presence of 4.3 M NaCl[7], while a temperature of around 50°C is sufficient to produce pure bassanite from solutions containing >6 M $CaCl_2$ and small amounts of $Na_2SO_4$ [13]. Tritschler et al. studied[8,9] the precipitation behaviour of calcium sulfate from mixtures of water and organic (co)solvents at different ratios and found that below a critical water content (40-50 wt%) bassanite forms as the main crystalline phase. These bassanite nucleation studies highlight the key role of hydration effects during the crystallization of calcium sulfate phases, which are likely to change substantially as the activity of water is significantly reduced at high salt/low water contents. Another observation pointing in the same direction is the formation of bassanite during the evaporation of droplets of $CaSO_4$ solutions at room temperature[15,16]. The relative humidity (RH < 80%) and the time-dependent availability of water appear to be the controlling factors for phase selection in this system.

Although the physicochemical conditions leading to bassanite formation in aqueous systems are relatively well established, distinct microscopic insights into the nucleation mechanisms and pathways are still lacking. To fill this gap, and to further improve our general model for calcium sulfate precipitation, we conducted in situ scattering measurements at small- and wide-angles (SAXS/WAXS) as well as complemented these with *in situ* Raman spectroscopic characterization, which allowed us to gather detailed information on the early stages of bassanite formation. Based on these experiments we show that the process of formation of bassanite from aqueous solutions is very similar to the formation of gypsum: it involves the aggregation of small primary species into larger aggregates, from which the crystalline phase develops. These data thus confirm our general model of $CaSO_4$ nucleation and provide clues to explain the abundant occurrence of bassanite on the surface of Mars (compared to its minimal presence in natural Earth surface environments).

**Methods**

*Synthesis of bassanite*

Calcium sulfate hemihydrate, $CaSO_4 \cdot 0.5H_2O$, *i.e.* bassanite, was synthesised by reacting equimolar high-salinity (4.3 M NaCl, >99 %, Sigma) aqueous solutions (deionized water, resistivity <18 MΩ·cm) of $CaCl_2 \cdot 2H_2O$ (pure, Sigma) and $Na_2SO_4$ (>99 %, Sigma) at 90°C, based on the following idealized reaction:

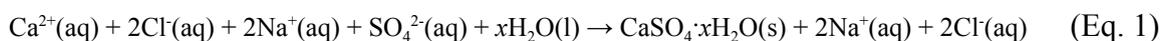

$Ca^{2+}(aq) + 2Cl^-(aq) + 2Na^+(aq) + SO_4^{2-}(aq) + xH_2O(l) \rightarrow CaSO_4 \cdot xH_2O(s) + 2Na^+(aq) + 2Cl^-(aq)$   (Eq. 1)

In Eq. 1, $x = 0.5$ in the case of bassanite. Prior to mixing, all solutions were equilibrated at 90°C. The high salinity and elevated temperature conditions reduce the activity of water, which promotes the precipitation of metastable hemihydrate, instead of the thermodynamically more stable calcium sulfate, $CaSO_4$, *i.e.* anhydrite[e.g. 7]. Bassanite precipitation was performed in a 200 mL temperature-stabilized glass reactor equipped with a reflux condenser to prevent any evaporation of water. The reaction temperature in the vessel was maintained by means of an oil bath on a hot plate, and a thermocouple was placed directly in the reactor to provide accurate feedback. The reacting solutions were continuously stirred at 350 rpm, and circulated through a custom-built PEEK flow-through cell with borosilicate glass capillary (ID 1.5 mm) using a peristaltic pump (Watson Marlow 120V/DV, flow ~4 mL/s). The connector tubes (total length of tubing of ID 2 mm was 1.5 m) used in the experiments were thermally insulated. The maximum temperature drop between the outlet and inlet of the reactor was <2°C at 90°C, as we recorded with a secondary in-line thermocouple.

Experiments started with 50 mL of a temperature-stabilized $CaCl_2$ aqueous solution inside the reactor. This solution was circulated through the capillary cell while the measurements (scattering or spectroscopy) were collected continuously as described below. $CaSO_4$ formation reactions were initiated through the injection of a corresponding volume of a temperature-stabilized $Na_2SO_4$ aqueous solution. Injection and mixing of the two solutions was achieved either with the use of a secondary peristaltic pump (Watson Marlow 120V/DV) as a remote fast injection system (scattering, mixing within 15 s) or manually (spectroscopy, mixing within 2 s).

*SAXS/WAXS data collection and processing*

In situ and time-resolved small- and wide-angle X-ray scattering (SAXS/WAXS) measurements of the synthesis of bassanite were carried out at beamline I22 of the Diamond Light Source Ltd (UK) (whereas, preliminary experiments were conducted at the NCD beamlime of the ALBA Synchrotron Light Facility and BM26 of the ESRF). Experiments were performed using a monochromatic X-ray beam at 12.4 keV aligned with a capillary of the flow-through setup, and scattered intensities were collected at small-angles with a Dectris Pilatus P3-2M, and at wide-angles with Dectris Pilatus P3-2M-DLS-L (2D pixel-array detectors). Transmission was measured by means of a photodiode installed in the beam-stop of the SAXS detector. The sample-to-detector distances allowed for a usable $q$-range of $\sim0.015 < q < \sim0.35$ Å$^{-1}$ in SAXS, and of $\sim0.17 < q < \sim6$ Å$^{-1}$ in WAXS. The scattering $q$-range at small-angles was calibrated against silver behenate and the corresponding measured intensity was calibrated to absolute units against glassy carbon. The $q$-range in WAXS was calibrated with a cerium(IV) oxide powder.

The recorded 2D scattering data from the SAXS and WAXS detectors were pre-processed using DAWN 2.11[17,18]. This pre-processing steps involved a number of corrections described by Pauw et al. in refs. [19,20] and implemented into DAWN, as well as integration to 1D scattering curves and subtraction of an instrumental background (i.e. an "empty beamline" background). The SAXS data were corrected for transmission and scaled to absolute intensity units. The measured WAXS could not be directly corrected for transmission because there was no photodiode installed for this detector. However, since the covered $q$-ranges of the SAXS and WAXS overlapped between 0.18 Å$^{-1}$ - 0.5 Å$^{-1}$, we scaled and matched the WAXS intensities against the corresponding corrected and normalised SAXS data, and as a result we obtained a continuous SAXS/WAXS raw scattering curves. These raw scattering curves were further corrected for a "solvent background" (see below).

The scattering data from the formation of $CaSO_4$ were acquired at 1 fps, with data collection starting when one solution started to be injected into the other. The injection period lasted 15 s. This created a dead-time, during which the reaction had already started, but the target physicochemical conditions were not yet reached. To consider this dead-time, we used a rolling average over a time series with a window of 15 s, and then the data were binned and averaged into increments of 3 s. We used an aqueous solution of 4.3 M NaCl at 90°C as a "solvent background", which was subtracted from the raw scattering curves. This background was measured at 1 fps over a period of 300 s, and averaged into a single scattering pattern of high signal-to-noise ratio. All these operations were performed automatically for all the SAXS-WAXS pairs, including the solvent backgrounds, using scripts written in Python/NumPy/Pandas[21,22].

*Raman measurements and data processing*

The spectroscopic measurements were performed using a HORIBA Jobin Yvon LabRAM HR800 VIS Raman microscope equipped with a 473 nm laser line, Olympus BXFM microscope with a motorized XYZ microscope stage, and a CCD detector (2048 x 512 pixel$^2$). The spectra were collected at full laser power through a 20x objective lens, at 40 s integration time and 3 accumulations, using a 100 μm entrance slit and an 800 μm pinhole aperture. For the Raman characterization we used exactly the same flow-through cell setup and synthesis conditions as for the scattering experiments. Before the start of an experiment the cell with an embedded capillary was firmly clamped onto the microscope stage, and the microscope was used to find the focal planes of the lower and upper walls of the capillary. In the next step the motorized stage was shifted by 0.5 mm in Z (1/2 of an ID of the capillary), so that the focal plane of the microscope was approximately in the middle of the capillary. After these initial adjustments, no other settings of the microscope and the spectrometer were changed to ensure that all collected spectra would be directly comparable with each other in terms of signal intensity. A series of spectra was collected *in situ* during the formation of bassanite ($CaSO_4 \cdot 0.5H_2O$) for a period of up to 60 min. In addition we measured an aqueous sulfate calibration series, for which we used dissolved $Na_2SO_4$ in 4.3 M NaCl at 90°C, at the concentrations of: 25, 29, 36, 42, 50, 62.5 and 100 mM. In addition, in a different set of similar experiments, we followed the formation of gypsum ($CaSO_4 \cdot 2H_2O$) at 21°C from a 50 mM calcium sulfate solution (SI: Fig. S1). We synthesised gypsum based on Eq. 1, but with $x$ = 2. In contrast to bassanite, gypsum was formed without any additional NaCl (at low-salinity conditions). These measurements involving gypsum were performed under different microscope settings than those at 90°C, and therefore the recorded absolute intensities are not directly comparable between the two sets of experiments.

For all the measurements, we monitored primarily the symmetric $v_1$ S-O stretching in the sulfate group (Raman shift at ~1000 cm$^{-1}$), which is the highest intensity band in any solid

CaSO$_4$ phase and in aqueous sulfates[23–27]. The exact position of the Raman shift depends on the sulfate environment. It changes readily for different sulfate-containing compounds and as such can be used for phase and compound identification[25,27–31]. In order to improve the time resolution the spectral acquisition range during the in situ measurements was limited to 1270 - 680 cm$^{-1}$. We used the area under the peak of the $v_1$ to determine the concentration of aqueous sulfate concentration[27] as a function of time, based on the calibration series. The collected spectra were corrected for a baseline using an Asymmetric Least-square Smoothing (ALS) algorithm[32], and the Raman bands were deconvoluted by non-linear fitting of Voigt functions using scripts written in Python/NumPy/Pandas[21,22].

**Theory**

The scattering patterns corresponding to the small-angle part of the data ($q < 1$ Å$^{-1}$) were fitted with the general expression for the intensity[33,34]:

$$I(q) = I_0 P(q) \cdot S(q) \qquad (Eq.\ 2)$$

In Eq. 2, $P(q)$ is the so-called form factor, which describes the shape and size of particles at mid- and high-$q$ ($q > 0.1$ Å$^{-1}$). Importantly the form factor fulfills the normalization condition so that $P(q\to 0) = 1$. In our fitting routine we tested several expressions for $P(q)$: a standard monodisperse sphere of radius $R$ [35]; a monodisperse cylinder of radius $R$ and length $L$ [33]; an approximated Guinier form factor of radius of gyration $R_g$ [36,37]; and a generalised approximated spherical form factor of a radius $R$ [38]. The detailed information on all these form factors can be found in the provided references. As we show later the actual choice of the form factor is of no significance, since the shape of the first formed, i.e. primary, particles cannot be determined, and hence the form factor only serves as a proxy to determine the volume and the radius of gyration of the primary species. By convention $S(q)$ is a structure factor function, which expresses how particles of the form factor $P(q)$ are arranged in space (and/or interact with each other), and therefore this function describes scattered intensity at low-$q$. The structure factor is normalised so that $S(q\to\infty) = 1$. For the actual $S(q)$ we used a previously derived expression for so-called "brick-in-the-wall" surface fractal aggregates[38] (Eq. 3):

$$S(q) = 1 + \frac{9\phi_b \Gamma(5-D_s)}{(2qr_0)^{6-D_s}} \frac{\sin[\pi(3-D_s)/2]}{3-D_s} \qquad (Eq.\ 3)$$

In Eq. 3, $\phi_b$ is a fraction of primary particles located at the fractal surface boundary from an overall pool of all the available primary particles (inside the aggregate or unaggregated ones in solution); $D_s$ is the surface fractal dimension (ranging from 2 for smooth surfaces, to 3 for very rough surfaces); and $r_0$, the primary particle radius. Based on the above normalizations of $P(q)$ and $S(q)$, the pre-factor $I_0 = (\Delta\rho)^2 NV^2$, where $\Delta\rho$ is the scattering length density contrast of the growing phase, $N$ is the number density of the primary species, and $V$ is the volume of the primary particle. Clearly, $V$ is dependent on equivalent size from $P(q)$, but the value of $I_0$ corresponds directly to a transitional intensity between the form factor and the structure factor so that $I_0 P(q\rightarrow 0) = I_0 S(q\rightarrow\infty)$. Therefore, we treated the pre-factor $I_0$ as an independent parameter in our fitting routines. The non-linear fitting of the data set was performed using the *optimize.least_squares* function which is part of the SciPy library[39]. The uncertainties of the fitted parameters were calculated from the jacobian matrix calculated by the *optimize.least_squares* function following ref.[40].

**Results and Discussion**

*Time resolved scattering during calcium sulfate precipitation*

Fig. 1 shows the evolution of the scattered intensity during bassanite precipitation from solution over a period of 1464 s. The merged measured intensities from the SAXS and WAXS detectors show a recorded signal that spans a $q$-range from ~0.015 Å$^{-1}$ to ~6 Å$^{-1}$. The coloured backgrounds indicate the $q$-ranges of the scattering curves collected with the SAXS (orange, Fig. 1) and WAXS (blue, Fig. 1) detectors. Considering three different length-scales, the following characteristic changes could be distinguished in the collected time-series (FIg. 1A):

(I) For 0.1 Å$^{-1}$ < $q$ < 1 Å$^{-1}$ the scattering intensity starts to decrease, and over a period of ~36 s a particle form factor develops. This form factor remains relatively unchanged until the end of the process. Characteristically, at 0.4 Å$^{-1}$ < $q$ < 1 Å$^{-1}$ the scattered intensity follows mostly a constant $I(q) \propto q^{-4}$ dependence, which implies the formation of an interface between the nano-sized particles and the surrounding medium (i.e. salty aqueous solution).

(II) At $q$ < 0.1 Å$^{-1}$ a gradual increase in intensity (up to ~100-fold) occurs, which also follows a $I(q) \propto q^{-4}$ dependence that starts to develop after ~36 s. This increase at low-$q$ implies the surface growth of large scattering features.

(III) For $q > 1$ Å$^{-1}$ changes inherent to the atomic scale are recorded. The initial decrease in intensity between 0 and 36 s, coincides with the changes discussed for $q$-range (I), and represents the evolution of broad scattering features from the medium matrix at 0 s, toward broad scattering features of a new phase different from the original medium. Most importantly, narrow diffraction peaks also emerge in this angular range after ~200 s, which indicates the formation of a crystalline material. These diffraction patterns match phase-pure bassanite, $CaSO_4 \cdot 0.5H_2O$, and no other crystalline phases were observed throughout the entire length of the experiment. In the next sections we present a more in-depth analysis of the processes taking place in ranges I and III.

*The evolution of the microstructural features at length scales I and II*

We interpret the observed evolution of the scattered intensities for $q < 1$ Å$^{-1}$ (Fig. 1A) in terms of the fast formation of nano-sized primary species (I), followed by their aggregation to micron-sized assemblies[3] resulting in a ~100-fold increase in intensity in range (II). The primary species reach a constant form factor with $I(q) \propto q^{-4}$, 36 s after the onset of the reaction. Consequently, different interfaces, particulate shapes, and/or polydispersity should be considered for the evolving form factor during the first 36 s. The decreasing scattering intensity in (I) for the initial 36 s can be explained in terms of decreasing scattering contrast as the primary species form relative to the background of 4.3 M NaCl solution. Such a scenario might involve e.g. formation of small anhydrous Ca-SO$_4$ clusters that develop a hydration sphere as they transition to larger primary units, and thus decrease their average electron density. The intensity increase in (II) concerns the aggregation of primary species after the aforementioned 36 s. Apart from several initial curves after the onset of aggregation, the intensity increase follows an approximate $I(q) \propto q^{-4}$ dependence (although the exponent is actually slightly less negative than -4, in particular in the later stages of the process, see below for further discussion). This ~$q^{-4}$ dependence implies the formation of surface fractal structures, and the overall intensity increase originates from the evolving structure factor ("brick-in-the-wall")[38]. Furthermore, the scattering curves for $q < 1$ Å$^{-1}$ show a characteristic transition plateau, at ~0.1 Å$^{-1}$, separating the form factor component (I) from the structure factor component (II).

Based on the above analyses, we fitted these scattering data using our "brick-in-the-wall" scattering model (Eq. 3) developed to describe gypsum precipitation from solution[38]. In this model the low-$q$ structure factor component in (II) corresponds to the surface fractal aggregates (Eq. 3), and the primary particles in (I) are described by a generic form factor for $P(q)$ (Eq. 2). Fig. 1B shows that the various form factors (spherical, cylindrical, etc.) describe in a similar way scattering for $q > 1$ Å$^{-1}$, but they do not reproduce the high-$q$ scattered intensity in (III). This form factor inadequacy is probably caused by the fact that we observe a transition in the probed length-scales. Eq. 2, which is generically used to describe the intensity at small angles, assumes a constant and homogeneous electron density contrast between the particles and the surrounding medium. When probing "inside" of the primary particles with an increasing scattering vector (the "probing yardstick"), this approximation becomes essentially too crude [e.g. 41] when reaching the atomic length-scales at $q > \sim 1$ Å$^{-1}$ in (III). Furthermore, at the mesoscale the shapes of the small clusters cannot be expected to adhere to simple geometries. Consequently, the actual form factor(s) of the primary particles cannot be represented by the simplified expressions for geometric objects, and for instance both cylindrical and spherical form factors fit the data equally well within a limited $q$-range (Fig. 1B). If we assume that the primary particles are indeed anisotropic and we fit with a cylindrical form factor (Fig. 1B), we obtain a radii, $R$ of ~4 Å and a length, $L$ of ~10 Å (and hence the radius of gyration of ~4 Å, from a classical mechanics relation, $R_g^2 = R^2/2 + L^2/12$). However, due to the apparent relatively low aspect ratio of this hypothetical cylinder's dimensions, the data do not exhibit any extensive $q$-range for which $I(q) \propto q^{-1}$ scaling is directly obvious for $q < 1$ Å$^{-1}$, as would be expected for rod-shaped objects of higher aspect ratios. Our bassanite data do not exclude an anisotropic shape (see Fig. 1B), but we cannot confirm or refute it unequivocally without e.g. performing a pair-distribution function analysis of the diffraction data and model fitting (which is not feasible for the current data set and measurement conditions). However, regardless of the used form factor expression, the as-obtained equivalent volumes and radii of gyration for the formed particles must be the same, because they represent meso-scale, and not atomic-scale parameters. Hence, the data were fitted with the generic Guinier form factor, which correctly approximates the size of the primary particles, ignoring any specific shape of the species. From these fits, we obtained several parameters (Fig. 2), which quantify the evolution of the system in length-scale ranges (I) and (II) over a period between the development of the form factor at 36 s, until the end of the experiment at 1464 s. The size of the primary particles is relatively constant after they finish forming at 36 s, although they do grow from $R_g$ ~4.05 to ~4.20 Å during the first 400 s

- 500 s of the processes (Fig. 2A). This is a minor, yet significant, change which indicates a change in size and/or polydispersity of the primary species, as they become part of the larger aggregates. This growth in $R_g$ coincides in time with a minor decrease (~ -8%) in $I_0$ from the initial ~0.0137 to ~0.0128 cm$^{-1}$ after 500 s (Fig. 2B). Importantly, the pre-factor normalised against the square of the primary particles' volume, $I_0/R_g^6$, shadows the changes in the original $I_0$. This is significant because $I_0/V^2 \propto I_0/R_g^6 \propto (\Delta\rho)^2 N$ (see the Theory section for the definitions). Hence, in the most plausible scenario where we assume that $\Delta\rho$ = constant, the change of $I_0/R_g^6$ will reflect the gradual decrease of the number density of the primary particles ($N$), which correlates negatively with the change in their size (the $R_g$). Consequently, this suggests that the primary particles undergo coalescence[42], prior to (i.e. it "kick starts") the crystallisation step (see below for further discussion).

Changes in $D_s$ (Fig. 2C) and $\phi_b$ (Fig. 2D) at longer length-scales reflect the evolution of the aggregates (composed of the primary species of a size expressed by $R_g$, Fig. 2A). $D_s$ and $\phi_b$ exhibit similar rates of change as those observed for $R_g$ and $I_0$ (Fig. 2A&B). Based on our definition of the "brick-in-the-wall" aggregates[38], $\phi_b$ reflects indirectly the size of the aggregates: it is the fraction of particles located at the surface of the aggregates, from a pool of particles within the aggregates and/or free unaggregated ones. Consequently, an increase in $\phi_b$ before 500 s translates into an increase in surface of the aggregates. As aggregation progresses the value of this parameter increases to reach a relatively stationary value of ~0.275% after 500 s. This can be best explained by the fact that the aggregates are formed from the free primary particles, which come together to form surface fractal larger objects. On the other hand, the $D_s$ initially assumes values of ~2 up to 500 s, which corresponds to an intensity scaling characteristic of smooth interfaces[43–45], but at this stage the obtained values in the trend exhibit high uncertainties, due to the limited $q$-range at which the structure factor manifests itself. Regardless, it seems that values of $D_s$ before 500 s are lower (smoother objects) than those after 500 s, when the value starts to increase to ~2.05, implying the roughening of the aggregates' surface. These trends can be explained taking into account two different models; coalescence occurs between the already formed aggregates[e.g. 46], or secondary nucleation occurring at the aggregates surfaces[e.g.47]. Both scenarios should lead to rougher objects, which most likely represent a transitory state that relaxes over time. This is probably reflected in the small, but continuous, decrease in $\phi_b$ in Fig. 2C after 500 s. Our simulations of surface fractal aggregates[38] showed that, counterintuitively, the internal coalescence of the aggregate does not decrease the fractal dimension (make a smoother object). Therefore, a simultaneous decrease in $\phi_b$ and increase in $D_s$ is best explained by small aggregates that coalesce with larger ones. They may e.g. close cavities of the larger aggregates, making them bulkier, while at the same time they are roughening at the outer surface e.g. in the course of the aforementioned secondary nucleation. Overall, it appears that the threshold of 500 s signifies an important change in the evolution of the nucleation process. The actual formation and growth of primary particles, as well as the nucleation of the "macroscopic" solid phase through the aggregation of these primary particles to larger structures, all occur before 500 s. Past the 500 s period, the primary particles no longer change, but an increase in $D_s$ (Fig. 2C) implies that aggregation between larger particles and/or surface induced nucleation and/or reorganisation of the aggregates is taking place.

*The evolution of the atomic-scale structural features at III*

For length-scale range (III), which corresponds mostly to the WAXS region of the scattering patterns, two major processes could be distinguished. At the earliest stages (<36 s) a disordered nanophase starts to emerge, indicated by scattering intensity increasing above the background signal (Fig. 3A). The time-scale and the sequence of these changes match the evolution of the form factor in (I) before 36 s. If we compare the scattering resulting from the forming nanophase with the aqueous media after 36 s of reaction, it becomes apparent that the nanophase is structurally different from the aqueous solvents, exhibiting three broad maxima at 1.89, 2.80 and 4.85 Å$^{-1}$ (Fig. 3B). The scattering from the nanophase in the patterns from the current bassanite experiments, shows a very high degree of similarity to the one we obtained for the precursor nanophase when we followed gypsum crystallization[4] (Fig. 3A). The major difference is in the position of the 2$^{nd}$ maximum, which in gypsum is shifted toward 3 Å$^{-1}$ (corresponding to a ~7% shorter *d*-spacing, Fig. 3A).

After 200 s of reaction, diffraction peaks start to appear (Fig. 4). The diffraction peaks were separated from the nanophase, using an Asymmetric Least-square Smoothing (ALS) routine and by treating the nanophase as a baseline (Fig. 4A). This analysis reveals that throughout the entire crystallisation, bassanite constituted the sole crystalline calcium sulfate phase (see inset in Fig. 4A). Noteworthy, this crystallisation process coincides with changes of the four parameters (Fig. 2) of the SAXS scattering model used to fit the curves in ranges I and II (in Fig. 1A), as well as scattering from the nanophase (range III in Fig. 1A). In essence, the scattering intensity of the form factor shifts marginally towards lower *q* (Fig. 4B), which implies a change in the shape and/or size of the primary species, which we measured as an increase in $R_g$ starting after ~200 s (Fig. 2). Similarly, the nanophase pattern in range III decreases in intensity simultaneously with the increasing intensity of the diffraction pattern (the inset in Fig. 4B and Fig. 4C), which suggests that the crystals grow at the expense of the nanophase. This "consumption" of the nanophase seems to be also associated with the decrease of the number density of the primary particles $N$ (from $I_0/V^2 \propto I_0/R_g^6 \propto (\Delta\rho)^2 N$ in Fig. 2B), which can be interpreted again as coalescence of the primary species. We quantified this transformation reaction (i.e. nanophase to crystalline bassanite) as a function of time using a scattering *pseudo*-invariant (i.e. *q*-limited invariant $Q = \int q^2 I(q) dq$ for 0.5 Å$^{-1}$ < q < 6 Å$^{-1}$) as a proxy for the contribution of each phase to the scattering profile (Fig. 4C). The overall decrease in the invariant of the nanophase amounted to ~20%, and the fastest rate of this decrease (down to 10%) during the initial 200 s of the processes, precedes the crystallisation of bassanite.

*In situ Raman spectroscopic characterization*

We cross-correlated the information derived from our SAXS/WAXS data with experiments where we followed in situ the evolution of vibrational bands in time resolved Raman spectra (symmetric S-O stretching $v_1$ mode) during the precipitation of CaSO$_4$ at 90°C (Fig. 5A). In CaSO$_4$ phases this is by far the strongest band, which is paramount for our experiments since we are dealing with very low volume fractions (φ << 1%) of solids in aqueous solutions. At the beginning of the reaction, after 120 s, a single intense band centred at 980.0±0.1 cm$^{-1}$ was present. This Raman shift is characteristic for $v_1$ of aqueous sulfate[27] (i.e., "free" sulfate). As the formation process continues, the intensity and the area under $v_1$ at 980 cm$^{-1}$ decreases, and progressively another band appears and grows at 1010±0.5 cm$^{-1}$ (Fig. 5A&B). The area under $v_1$ at 980 cm$^{-1}$ is proportional to the concentration of sulfate species in solution (Fig. 5C and ref. [27]) and can be used to monitor quantitatively the consumption of sulfate from solution during the precipitation process (Fig.5B&C). The band at 1010 cm$^{-1}$ corresponds to the $v_1$ mode of SO$_4^{2-}$ in bassanite[25]. The observed position of the band is redshifted in comparison with the published reference data for bassanite (typically[25] 1014-1015 cm$^{-1}$), and is relatively close to the reported position of $v_1$ in gypsum[25] (~1008 cm$^{-1}$). Thus, we also used our experimental Raman setup to perform *in situ* synthesis of gypsum (see Methods and SI: Fig. S1) and found that gypsum exhibits the band at 1006.8±0.1 cm$^{-1}$ (and $v_1$ of liquid sulfate also at 980.0±0.1 cm$^{-1}$). Hence, there is a discrepancy of >1 cm$^{-1}$ between our measurements and those reported in literature (1008 cm$^{-1}$), but the position of $v_1$ in aqueous sulfate is the same for both data sets and matches that reported in the literature[e.g. 27]. Hence, the $v_1$ band of gypsum is redshifted due to plausible local structural differences between as-formed gypsum *in situ* in solution, and (highly) crystalline dry references used in other studies[25 and the references therein]. The position of $v_1$ band in bassanite is even further redshifted (~5 cm$^{-1}$) with respect to the reference data likely for the same reason (particularly as synthetic bassanite used for Raman is usually dehydrated gypsum and not like in our case bassanite formed directly from high ionic strength solutions). Therefore, it is possible that the shift in the position of the band originates e.g. from the local disorder/defects in the structure of bassanite, which is corroborated by the persistent presence of the disordered nanophase (Fig. 4C). Indeed, similar strong systematic redshifting was observed e.g. for $Mn_3O_4$ nanocrystals[48] of various sizes, and in this regard our earlier work on mesocrystallinity in calcium sulfate showed that single crystals of bassanite were persistently composed of smaller particle-like domains just 10-20 nm in size[10].

*The kinetics of bassanite precipitation*

To extract a kinetic proxy for the different steps of the precipitation reaction we compared the normalised changes as a function of time for three parameters (Fig. 5D): (i) the *pseudo*-invariant of the crystalline phase from WAXS (Fig. 4C), (ii) the fraction of primary particles located at the fractal surface boundary, $\Phi_b$ (Fig. 2D) from SAXS, (iii) and the integral area of the $v_1$ band at 1010 cm$^{-1}$ from Raman spectroscopy (Fig. 5A). From scattering data, the first trend corresponds directly to the crystallisation rate, whereas the second represents the overall growth rate and aggregation of particles. These two parameters are not identical: the crystallization rate represents merely the formation of bassanite, whereas the surface fractal fraction $\Phi_b$ is a proxy for the overall formation, growth and/or aggregation of any solid phase regardless of its crystallinity. Fig. 5D highlights that the formation of the solid material occurs significantly faster than the actual crystallization process, which indicates that bassanite is formed within/from the preceding solid precursor nanophase. The rate of evolution of bassanite in Raman matches closely the two proxies from scattering. Noteworthy, the initial growth rate from Raman spectra follows the kinetics of the changes in SAXS, and after ~500 s it follows the crystallisation trend. As discussed above (Fig. 2), the threshold of 500 s appears to mark a change in the growth processes, most likely nucleation of the solid phase through an aggregation of primary particles, and a subsequent transition to restructuring or coalescence of the primary particles within the larger aggregates. Hence, the kinetic trend from the Raman measurements also implies that what we observe as a nanophase and a crystalline bassanite phase in scattering, may in fact be similar to each other in terms of the local structure.

*Gypsum versus bassanite: a common precipitation pathway*

Based on our previous (also in situ) measurements of the precipitation pathway of gypsum and the outcome of $CaSO_4$ precipitation experiments, we hypothesized that all calcium sulfate phases could share a common precipitation pathway[1,2]. The in situ data of the precipitation pathway of bassanite presented in this work, allow us to compare the crystallization process of bassanite with that of gypsum. In a nutshell, gypsum formation can be summarized as follows:

1. First sub-3 nm primary species[3] appear in a solution supersaturated with respect to gypsum. These particles are composed of anhydrous Ca-$SO_4$ cores[3,4] that almost certainly contain water coordinated to their surfaces. Their diameter is defined by a single Ca - Ca / Ca - S distance and equals ~ 5 Å. The aggregation of these particles leads to the formation of a disordered precursor phase[4] of the "brick-in-the-wall"-type[38], i.e. they form internally dense surface fractal aggregates (yielding characteristic scattering patterns).

2. Creation of a crystalline lattice, i.e. gypsum crystallization, occurs through the reorganization of the aggregates and leads to imperfect mesocrystals, which preserve an imprint of the particle-mediated crystallisation process[10].

3. The disordered phase persists in coexistence with the crystalline phase[49,50], constituting an important part of the solids in solution after the crystallisation process has essentially come to a halt[4].

Taking into account the above summary, the processes of bassanite precipitation from solution is in fact comparable to that of gypsum (Fig. 6):

1. Bassanite formation from aqueous ions also starts with the appearance of nanosized primary species. The scattering data imply that these species have defined shapes, but they cannot be directly assigned to any simple geometry (e.g. rods). The primary species are overall smaller ($R_g \sim 4$ Å) than those we observed for gypsum at 21 °C and low salinity[3], where particles were found to be >27 Å in length and ~2.5 Å in radius, yielding $R_g > 8$ Å.

2. The obtained scattering data were best fitted using our "brick-in-the-wall" scattering model (Eq. 3) developed to describe gypsum precipitation from solution[38], thus confirming that the primary species leading to bassanite formation also aggregate to form "brick-in-the-wall" aggregates. The observed disordered phase is still remarkably analogous in structure and behaviour to what we observed for gypsum. Also similar to gypsum, this amorphous phase prevails throughout the entire growth and crystallisation processes (see Fig. 1 and ref. [4]), and constitutes a majority phase (Fig. 4C). The observed difference in the structure of disordered phases of bassanite and gypsum (Fig. 3) bears certain conceptual resemblance with the protostructures reported for amorphous calcium carbonate (ACC), which lead to the crystallization of the distinct calcium carbonate phases[51–55].

3. The crystallisation occurs through the formation and reorganization of the aggregates, and potentially involves the coalescence of smaller units. Most likely, this leads to the formation of mesostructured single bassanite crystals, as we previously reported using detailed TEM characterization[10].

In overall, the nucleation and growth mechanism proposed in Fig. 6 adheres to the novel concepts of processes leading to the formation of minerals from ions in aqueous solutions, developed in recent years [e.g. 56–58]. In this regard, the original, and rather naive, 'textbook' image of these phenomena, stemming from the adaptation of classical one step nucleation and growth theories, has increased in complexity (i.e. multistep nucleation) due to the discovery of a variety of precursor and intermediate species. These include solute clusters (e.g. prenucleation clusters, PNCs[59]), liquid(-like) phases, as well as amorphous and nanocrystalline solids etc.

**Conclusions**

The formation process of bassanite from aqueous solutions is very much alike to that of gypsum. Both processes involve the nucleation and aggregation of nanosized primary species into large "brick-in-the-wall" structures, from which crystals emerge. These "brick-in-the-wall" aggregates constitute a disordered precursor nanophase. There is however a difference in the structure of the disordered nanophase observed for bassanite vs gypsum. This is likely a result of the fact that bassanite is the 'stable' $CaSO_4$ phase at high temperature and/or salinity (low water activity), while gypsum is stable at low temperatures and lower ionic strengths. Despite all evidence of these differences, the thermodynamics/kinetics of these two $CaSO_4$ phases and any inter-transformations between them is still to be explored in future research.

From a thermodynamic point of view bassanite is a metastable phase of calcium sulfate, which in the Earth surface environment is only found in minor quantities, usually as a product of a partial dehydration of gypsum. Nonetheless, through this, and preceding [7–10], works it is now fully demonstrated that under conditions promoting (very) low water activity bassanite can form directly from solution at relatively low temperatures and can persist for long times in an aqueous environment. This information is key to help unravel the environmental conditions that lead to stable/persistent bassanite formations observed on the (sub)surface of Mars[1,60–63]. Previous research has postulated that during certain time periods extreme high salinity solutions (brines) were present on the surface of Mars[e.g 64,65]. Under those conditions the precipitation of bassanite, even at relatively low temperatures, is plausible, which would provide a possible route to explain the significant bassanite deposits on Mars.

**Acknowledgments**

This research was supported by a Marie Curie grant from the European Commission in the framework of the NanoSiAl Individual Fellowship (Project No. 703015) to T.M.S. The GFZ team was also generously funded by the Helmholtz Recruiting Initiative (Award No. I-044-16-01) to L.G.B. We thank Diamond Light Source for access to beamline I22 through grant number SM16256-1. We thank Dr. Andrew Smith from I22 for his technical support during our beamtime at I22. Also the preliminary SAXS experiments conducted at the NCD beamline of the ALBA Synchrotron Light Facility (Spain) and BM26 beamline of the ESRF (France), through grant numbers 2014071007 and 26-02-735, respectively, are acknowledged.

**Supporting Information Contents**

Figure S1. *In situ* Raman spectrum after 25 mins from 50 mM $CaSO_4$ at 21°C.

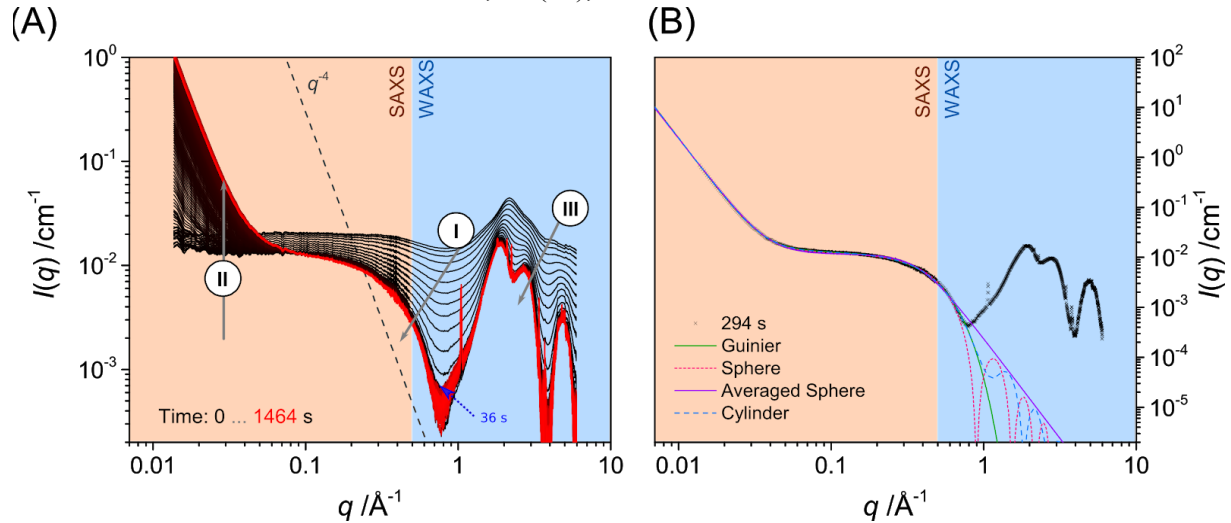

Figure 1. A) *In situ* and time-resolved scattering intensities collected from 100 mM $CaSO_4$ in 4.3 M NaCl at 90°C, using two detectors at different angular ranges: SAXS - the orange background, and WAXS - the blue background. Each curve corresponds to a time step of 3 s. The characteristic length-scales are marked: (I) the mid-$q$ evolution of a nano-particulate form factor; (II) the low-$q$ growth of large objects; (III) the high-$q$ evolution of atomic-scale features and the diffraction peaks. The dashed line indicates a Porod scattering profile. The gray arrows indicate the direction of the considered changes; B) a selected data set at 294 s, which shows the features of the fitting model, as described in the Scattering Model section; the low-$q$ part < 0.1 Å$^{-1}$ (II in A), is described predominantly by the "brick-in-the-wall" surface fractal structure factor from Eq. 3, where $\phi_b$ = 0.139±0.001%, $D_s$ = 2±0.006, $r_0 = (5/3)^{1/2}R_g$ in which
$R_g$ = 4.188±0.003 Å, and from Eq. 2, pre-factor $I_0$ = 0.01319±0.00001 cm$^{-1}$. The 0.1 Å$^{-1}$ < $q$ < 1 Å$^{-1}$ range (I in A) is best fitted by any generic form factor of $R_g$ = 4.188±0.003 Å, where the specific forms of $P(q)$ yield similar fits (a sphere, a cylinder, an averaged sphere, and a Guinier approximated form factor, see the main text for references). A better differentiation was not feasible because the part of the scattering intensity, which would allow us to distinguish among these different shapes, was dominated by scattering from the disordered phase (III in A). Similar fits were obtained for all the scattering curves from 45 s to 1464 s (see Fig. 2).

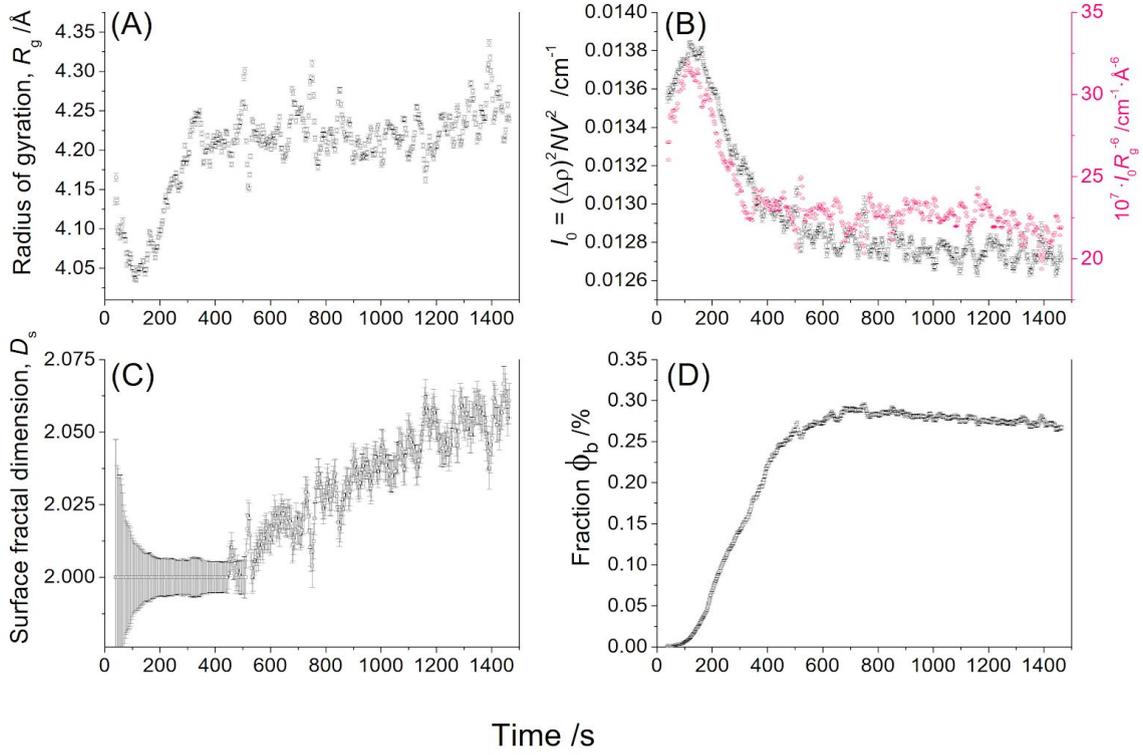

Figure. 2. The evolution of the four parameters characterising a scattering model (Eqs. 1 and 2) used to quantify the patterns in Fig. 1A for ranges (I) and (II) between 39 s and 1464 s: A) the radius of gyration. $R_g$; B) the pre-factor $I_0 = (\Delta\rho)^2 NV^2$ and the derived volume-normalised $I_0/R_g^6 \propto (\Delta\rho)^2 N$; C) the surface fractal dimension $D_s$; and D) the fraction of primary particles located at the fractal surface boundary, $\phi_b$.

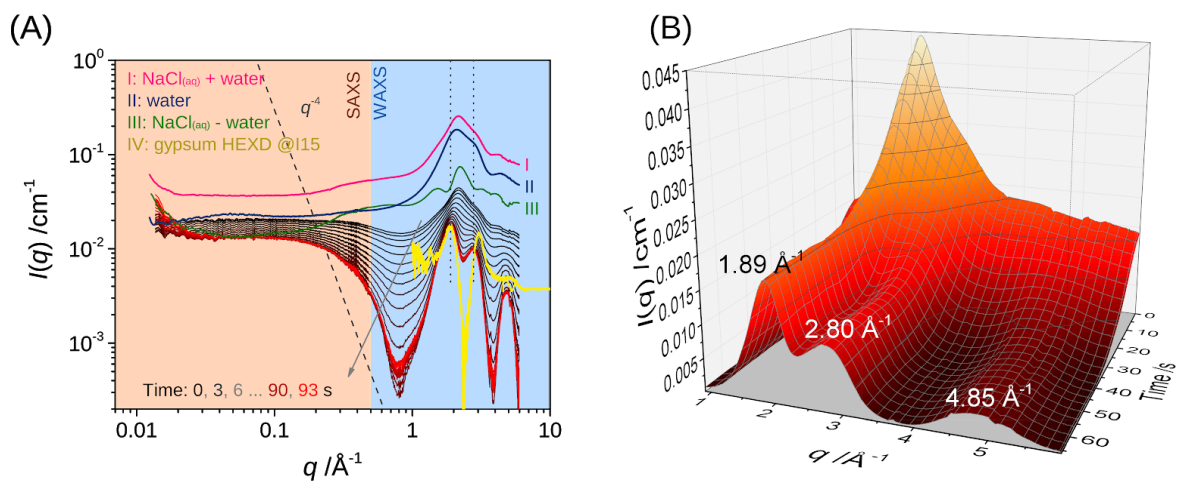

Figure 3. A) The early stages of the formation of bassanite, showing the transformation from the solvent/medium dominated scattering pattern at 0 s to a new disordered phase at 93 s. The as-formed disordered phase is structurally different from: (I) a 4.3 M aqueous solution of NaCl at 90°C, (II) pure water at 90°C, or (III) the difference between I and II. The dotted vertical lines highlight the two major maxima positions of the disordered phase, which are different from the solvents in I-III. The high-energy X-ray diffraction (HEXD) pattern in IV was measured for the reaction of formation of gypsum, and originates from our earlier work[4]. B) The time-resolved WAXS patterns from the disordered phase show the evolution of the three broad maxima at 1.89 Å$^{-1}$, 2.80 Å$^{-1}$, 4.85 Å$^{-1}$.

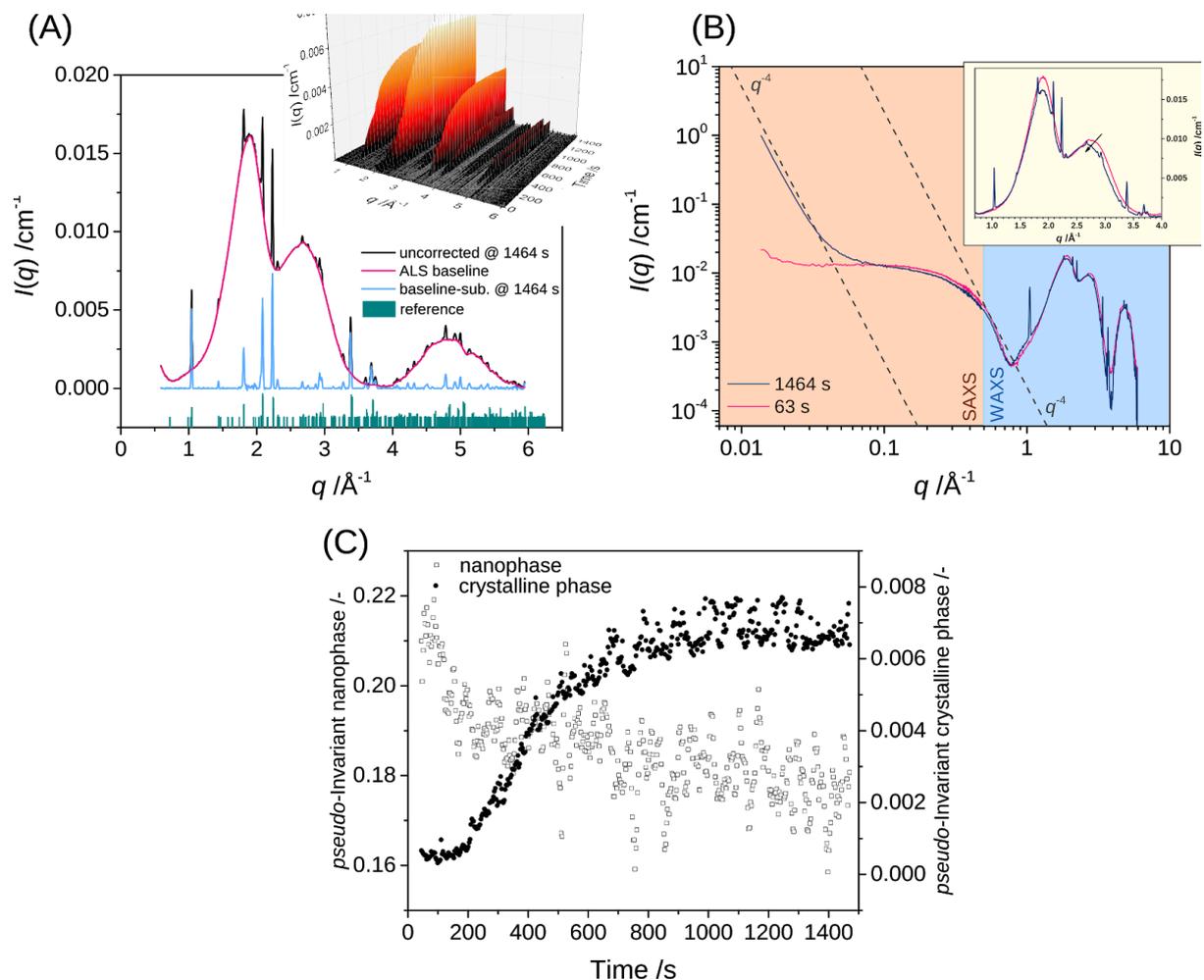

Figure 4. A) The diffraction pattern at 1464 s corresponds to a phase-pure bassanite[61]. The diffraction pattern was separated from the scattering from the nanophase by applying an ALS baseline correction[32]. The inset shows that bassanite was a sole crystalline phase throughout the entire measurement, and confirms that diffraction peaks started to appear after at least 105 s of the processes; B) The figure highlights very small changes in the scattering pattern for $q > 1$ Å$^{-1}$ (ranges I and III from Fig. 1), which include a small shift of the form factor at (I) (see Fig. 1) towards lower $q$, and a very small decrease of the overall intensity of the nanophase (the inset); C) The relative scattering pseudo-invariant as a function of time corresponding to the nanophase and the crystalline phase.

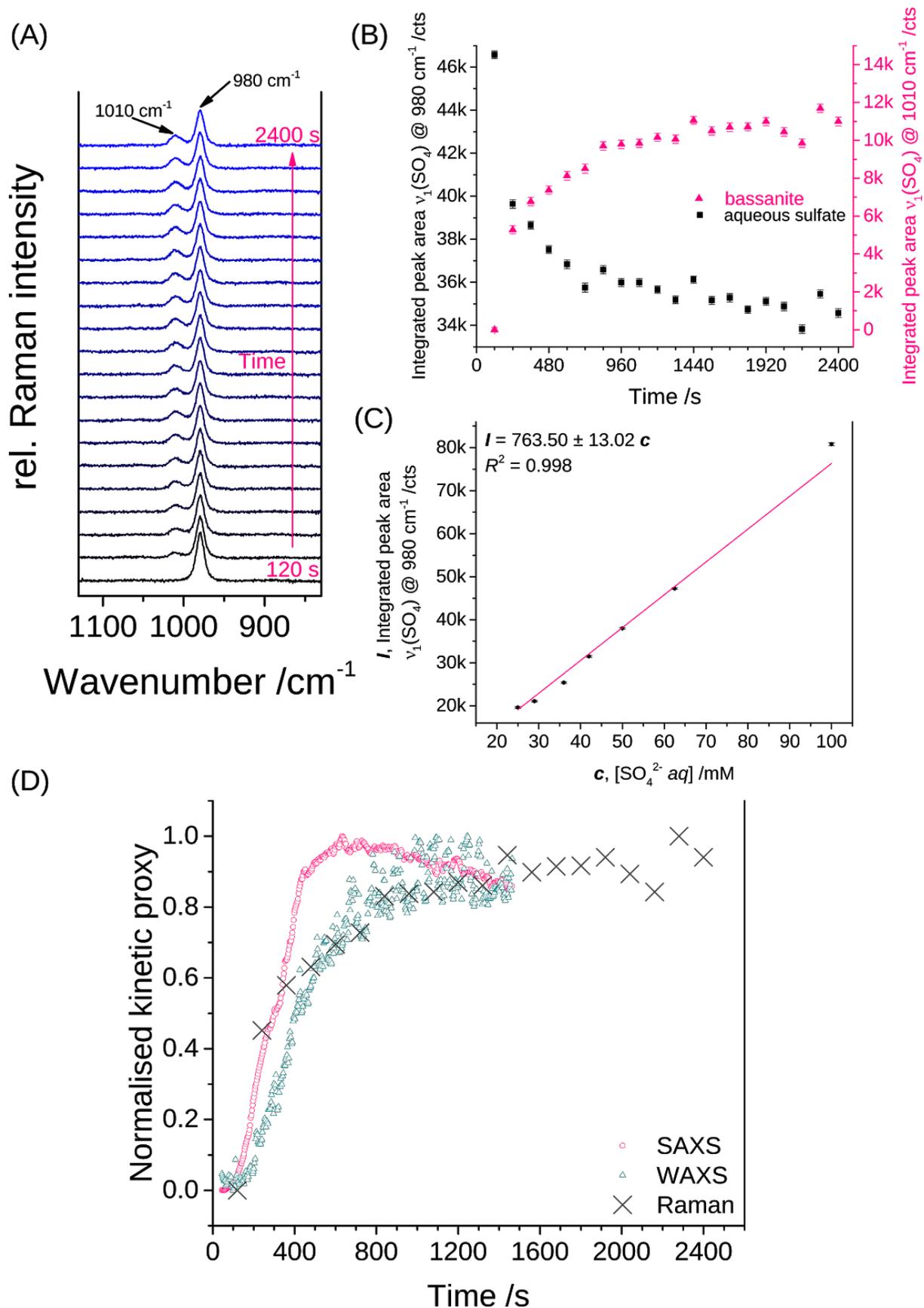

Figure 5. A) *In situ* Raman spectra from the processes of precipitation of $CaSO_4$ in 4.3 M NaCl at 90°C. The strong band at 980 cm$^{-1}$ corresponds to $v_1$ mode of aqueous sulfate, whereas the peak at 1010 cm$^{-1}$ is a $v_1$ band of the growing solid phase. Each spectrum corresponds to a time step of 120 s. All spectra were baseline-corrected and a constant offset in intensity was introduced for clarity; B) Integrated areas of both $v_1$ bands from (A) as a function of time; C) Integrated area of the $v_1$ band in aqueous sulfate from $Na_2SO_4$ in 4.3 M NaCl at 90°C, as a function of its concentration; D) Comparative plot of reaction progression from the small- and wide-angle scattering data as well as the Raman pattern showing correspondence between kinetic pathways (see main text for an explanation of the kinetic proxies).

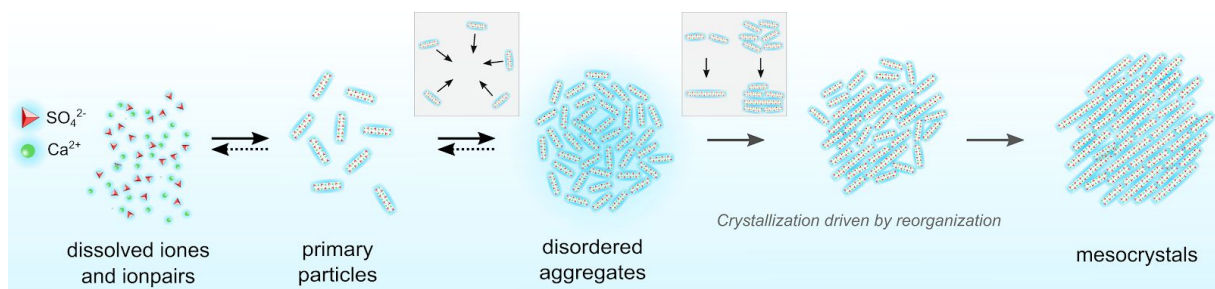

Figure 6. Generic model of the steps leading to the formation of crystalline CaSO$_4$ phases based on the results of in situ scattering and vibrational spectroscopic experiments. After the onset of supersaturation ions and ion pairs join to form primary species of CaSO$_4$ (~1 to 3 nm in length). With time these primary species assemble into larger (>100 nm) disordered surface-fractal aggregates. Reorganization of the primary species within those disordered aggregates results in the formation of CaSO$_4$ mesocrystals. The selection of the dominant crystalline phase essentially depends on the water activity of the medium, which is modulated through e.g. temperature and salinity of the solution. The insets show the successive evolution at the nanoscale, where the primary units first aggregate and then grow, by coalescence, yielding structural domains of increasing size and crystallinity. Noteworthy, even in the final crystalline end product the ordered domains remain separated by partially disordered domains.

# TOC graphic

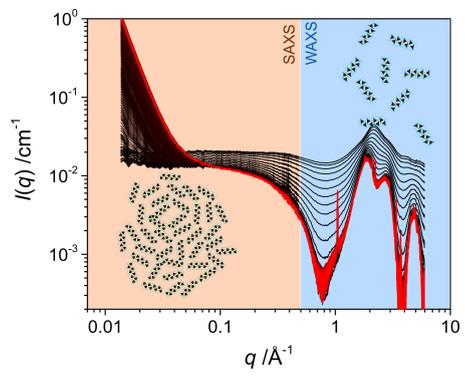

# Supporting Information

To

**Nucleation pathway of calcium sulfate hemihydrate (bassanite) from solution: implications for calcium sulfates on Mars**

Tomasz M. Stawski*[ab], Rogier Besselink[ac], Konstantinos Chatzipanagis[ad], Jörn Hövelmann[ae], Liane G. Benning[af], Alexander E. S. Van Driessche*[c]

* - corresponding author(s):

tomasz.stawski@bam.de; alexander.van-driessche@univ-grenoble-alpes.fr

[a]German Research Centre for Geosciences, GFZ, Interface Geochemistry, Telegrafenberg, 14473 Potsdam, Germany;

[b]Federal Institute for Materials Research and Testing (BAM), 12205 Berlin, Germany;

[c]Université Grenoble Alpes, Université Savoie Mont Blanc, CNRS, IRD, IFSTTAR, ISTerre, F-38000 Grenoble, France;

[d]Theoretical and Physical Chemistry Institute, National Hellenic Research Foundation, 48 Vassileos Constantinou Ave., 116 35 Athens, Greece;

[e]Remondis Production GmbH, Brunnenstraße 138, 44536, Lünen, Germany;

[f]Department of Earth Sciences, Freie Universität Berlin, 12249 Berlin, Germany.

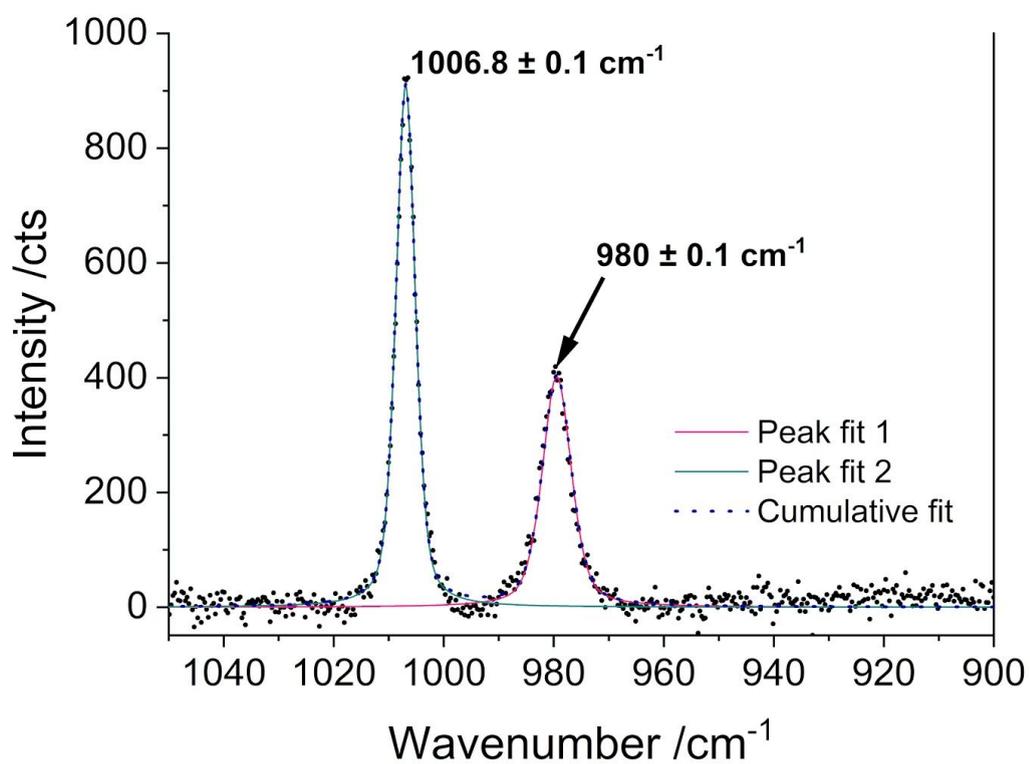

Figure S1. *In situ* Raman spectrum after 25 mins from 50 mM $CaSO_4$ at 21°C. The strong band at 980 $cm^{-1}$ corresponds to $v_1$ mode of aqueous sulfate, whereas the peak at 1006.8 $cm^{-1}$ is a $v_1$ band of the growing crystalline gypsum phase. The spectrum is baseline-corrected and the peaks are fitted with a Voigt function (also see Methods).